\documentclass[runningheads]{llncs}

\usepackage[english]{babel}
\usepackage[utf8]{inputenc}
\usepackage[T1]{fontenc}
\usepackage[pdftex]{hyperref}

\usepackage{amssymb}
\usepackage{amsmath}

\usepackage{amsthm}
\usepackage{amsfonts}
\usepackage{array}

\usepackage{multirow}
\usepackage{adjustbox} 
\usepackage{stfloats} 
\usepackage[caption=false]{subfig} 
\usepackage{booktabs} 
\usepackage{url}
\usepackage{wrapfig} 
\usepackage{graphicx} 
\usepackage{textcomp}

\usepackage[colorinlistoftodos]{todonotes} 
\usepackage{xcolor} 
\def\BibTeX{{\rm B\kern-.05em{\sc i\kern-.025em b}\kern-.08em 
    T\kern-.1667em\lower.7ex\hbox{E}\kern-.125emX}}
\usepackage{colortbl} 
\usepackage[misc]{ifsym} 

\hypersetup{pdfborder=0 0 0} 
\newcommand{\sectionname}{Sec.}
\newcommand{\colorcell}{ \cellcolor{gray!35} } 
\mathchardef\mhyphen="2D 
\newcolumntype{L}[1]{>{\raggedright\arraybackslash}p{#1}}
\newcolumntype{C}[1]{>{\centering\arraybackslash}p{#1}}
\newcolumntype{R}[1]{>{\raggedleft\arraybackslash}p{#1}}

\usepackage{paralist}
\theoremstyle{definition}
\usepackage{comment}
\newtheorem{defn}{Definition}[]

\makeatletter
\def\th@definition{
  \thm@notefont{}
  \normalfont 
}
\makeatother

\begin{document}

    \graphicspath{{figures/}} 


\title{Can I Trust My Simulation Model?\\Measuring the Quality of Business Process Simulation Models\thanks{Work funded by the European Research Council (PIX Project) and the National Science and Engineering Research Council (NSERC) grants held by Opher Baron, Dmitry Krass, and Arik Senderovich.}}

\titlerunning{Measuring the Quality of Business Process Simulation Models}

\author{
  David Chapela-Campa\inst{1}\textsuperscript{(\Letter)} \and 
  Ismail Benchekroun\inst{2} \and 
  Opher Baron\inst{3} \and \\
  Marlon Dumas\inst{1} \and 
  Dmitry Krass\inst{3} \and 
  Arik Senderovich\inst{4}}
\institute{
  University of Tartu \\
  \email{\{david.chapela, marlon.dumas\}@ut.ee} \and
  Department of Statistics, Faculty of Arts \& Science, University of Toronto\\
  \email{ismail.benchekroun@mail.utoronto.ca} \and
    Rotman School of Management, University of Toronto\\
  \email{\{opher.baron, dmitry.krass\}@rotman.utoronto.ca} \and
  School of Information Technology, York University\\
  \email{sariks@yorku.ca}
}

\authorrunning{D. Chapela-Campa et al.}

\maketitle

\begin{abstract}
Business Process Simulation (BPS) is an approach to analyze the performance of business processes under different scenarios. For example, BPS allows us to estimate what would be the cycle time of a process if one or more resources became unavailable. The starting point of BPS is a process model annotated with simulation parameters (a BPS model). BPS models may be manually designed, based on information collected from stakeholders and empirical observations, or automatically discovered from execution data. Regardless of its origin, a key question when using a BPS model is how to assess its quality. In this paper, we propose a collection of measures to evaluate the quality of a BPS model w.r.t.\ its ability to replicate the observed behavior of the process. We advocate an approach whereby different measures tackle different process perspectives. We evaluate the ability of the proposed measures to discern the impact of modifications to a BPS model, and their ability to uncover the relative strengths and weaknesses of two approaches for automated discovery of BPS models. The evaluation shows that the measures not only capture how close a BPS model is to the observed behavior, but they also help us to identify sources of discrepancies. 

\end{abstract}

\setcounter{footnote}{0} 


\section{Introduction\label{sec:introduction}}

Business Process Simulation (BPS) is a technique for estimating the performance of business processes under different scenarios~\cite{DBLP:books/sp/DumasRMR18}. BPS enables analysts to address questions such as ``what would be the cycle time of a process if one or more resources became unavailable?'' or ``what would be the impact of automating an activity on the waiting times of other activities in the process?''.
The starting point of BPS is a process model, e.g.\ in the Business Process Model and Notation (BPMN)\footnote{\url{https://www.bpmn.org/}}, enhanced with simulation parameters~\cite{DBLP:journals/bise/RosenthalTS21} (herein, a BPS model). These simulation parameters capture, for example, the processing times of each activity or the rate at which new process instances (cases) are created.

BPS models may be manually created based on information collected via interviews or empirical observations, or they may be automatically discovered from execution data recorded in process-aware information systems (event logs)~\cite{DBLP:journals/dss/CamargoDG20,DBLP:journals/peerj-cs/CamargoDR21,DBLP:journals/bise/MartinDC16,DBLP:journals/is/RozinatMSA09}. Regardless of the origin, a key question when using a BPS model is how to assess its quality. This question is particularly relevant when tuning the simulation parameters. Several approaches have been proposed to address this problem. However, these approaches are either manual and qualitative~\cite{DBLP:journals/is/RozinatMSA09} or they produce a single number that does not allow one to identify the source(s) of deviations between the BPS model and the observed reality~\cite{DBLP:journals/dss/CamargoDG20,DBLP:conf/caise/FraccaLAT22}.

In this paper, we study the problem of automatically measuring the quality of a BPS model w.r.t.\ its ability to replicate the observed behavior of a process as recorded in an event log. We advocate a multi-perspective approach to this problem, thus proposing a set of quality measures that address different perspectives of process performance. The starting point is the idea that a good BPS model is one that generates traces consisting of events similar to the observed data. Accordingly, the proposed approach maps an event log produced by the BPS model and an event log recording the observed behavior into  histograms or time series capturing a given perspective, and then compares the resulting histograms or time series using a distance metrics.

We conduct a two-fold evaluation of the measures using synthetic and real-life datasets.
In the synthetic evaluation, we study the ability of the proposed measures to discern the impact of modifications to a BPS model,
whereas in the real-life evaluation, we analyze their ability to uncover the relative strengths and weaknesses of two approaches for automated discovery of BPS models.


The rest of the paper is structured as follows.
\sectionname~\ref{sec:background} gives an overview of prior research related to the discovery and evaluation of BPS models.
\sectionname~\ref{sec:preliminaries} introduces relevant process mining concepts and distance measures.
\sectionname~\ref{sec:approach} analyzes the problem and proposes a set of measures of quality of BPS models.
\sectionname~\ref{sec:evaluation} discusses the empirical evaluation, and \sectionname~\ref{sec:conclusions} draws conclusions and sketches future work.


\section{Background\label{sec:background}}

\subsection{Business Process Simulation Models}

A BPS model consists of a stochastic control-flow (CF) model, an activity performance model, and an arrival and congestion model.
The stochastic model is composed of a process model (e.g., a BPMN model or a Petri net) and a stochastic component capturing the probability of occurrence of each path in the model (branching probabilities). In a BPS model, the  stochastic model is enhanced by adding an activity performance model, which determines the duration of the  activity instances (e.g., by associating a parametric distribution to each activity in the model).
Finally, in a BPS model, an arrival and congestion model determines when new cases arrive in the system, and when the execution of each enabled activity instance starts, given the available resource capacity.

Traditionally, BPS models are  constructed manually by experts. Recent approaches advocate for the automated discovery of BPS models from event logs.
Below, we consider two such approaches. The first one, namely \emph{SIMOD}~\cite{DBLP:journals/dss/CamargoDG20}, starts by constructing a stochastic process model by applying the SplitMiner algorithm~\cite{DBLP:journals/kais/AugustoCDRP19} to discover a BPMN model from the input log, and replaying the traces of the log to calculate the branching probabilities. Next, SIMOD discovers the activity performance model (activity duration distributions) and a congestion model consisting of: \textit{i)} a case inter-arrival time distribution; \textit{ii)} a set of resources, their availability timetables, and the activities they perform; and \textit{iii)} the distribution of extraneous waiting times between activities (i.e.\ waiting times not attributable to congestion)~\cite{DBLP:conf/icpm/Chapela-CampaD22}. Once a BPS model is discovered, its parameters are tuned to fit the data using a Bayesian hyper-parameter optimizer.


The second BPS model discovery technique we consider is \emph{ServiceMiner}$^{\copyright}$.
ServiceMiner operates in three steps:
\textit{i)} data preprocessing, where techniques for data cleaning and categorical feature encoding are applied;
\textit{ii)} data enhancement, where new data attributes that capture trend, seasonality, and system congestion are created using methods described in~\cite{DBLP:conf/aaai/SenderovichBGW19};
and \textit{iii)} model learning, where the BPS model is created by combining process discovery, queue mining (learning of queueing building blocks from data), and machine learning (to boost the accuracy of arrival and activity time generation).
For process discovery, ServiceMiner mines a Markov chain, estimating the case routing probabilities between consecutive activity pairs.
An abstraction mechanism allows for filtering out rare activities, paths, and transitions.
Next, using queue mining, the various queueing building blocks are fitted from data, by using techniques described in~\cite{DBLP:journals/is/SenderovichWYGM16}.
Lastly, ServiceMiner applies a machine learning technique that uses congestion features that come from queueing theory, which, via cross-validation, leads to accuracy improvements when generating inter-arrival times and activity durations.

While the evaluation reported below focuses on BPS models discovered by SIMOD and ServiceMiner, the proposed measures can be used to assess the quality of any model that generates event logs. For example, the proposal can also be used to evaluate generative deep learning models of business processes~\cite{DBLP:journals/peerj-cs/CamargoDR21}. On the other hand, it cannot be used to assess coarse-grained BPS models, e.g.\ based on system dynamics~\cite{DBLP:journals/access/PourbafraniA22}, unless these are refined to generate event logs.



\subsection{Quality Measures for Business Process Simulation Models}


Leemans et al.~\cite{DBLP:conf/caise/LeemansP20,DBLP:conf/bpm/LeemansSA19} and Burke et al.~\cite{DBLP:conf/icpm/BurkeLWAH22} studied the evaluation of stochastic models using, among other measures, Earth Movers' Distance.
However, they focus solely on the control-flow perspective, and their 
purpose is mainly conformance checking.
In this paper, we focus on the assessment of BPS model quality considering both temporal and control-flow dimensions.


Prior studies have considered the evaluation of BPS models.
Rozinat et al.~\cite{DBLP:journals/is/RozinatMSA09} perform an evaluation of BPS models following manual comparisons. However, they do not propose concrete and automatable evaluation measures.
Camargo et al.~\cite{DBLP:journals/peerj-cs/CamargoDR21} study the performance of data-driven simulation and deep learning techniques, proposing measures that combine the control-flow and the temporal perspectives.
The latter measures are not scalable, and they do not identify the sources of discrepancies between BPS models. To overcome these shortcomings, we propose an approach that views the process from different perspectives and provides a separation of concerns between the three BPS model components (the stochastic model, the activity performance model, and the congestion model). We then propose efficient measures for each component.


\section{Preliminaries\label{sec:preliminaries}}




\subsection{Event Logs\label{sec:event_log}}

Modern enterprise systems maintain records of business process executions, which can be used to extract \emph{event logs}: sets of timestamped events capturing the execution of the activities in a process~\cite{DBLP:books/sp/DumasRMR18}.
We assume that each event record in the log relates a case, an activity, and an 
activity start and end timestamp (as in Table~\ref{tab:example_log}). 
We shall refer to events and activity instances interchangeably, even though they could mean different things in other contexts.
Let $\mathcal{E}$ be the universe of events, $C$ be the universe of case identifiers, $A$ be the set of possible activity labels, and $T$ be the time domain. 
\begin{defn} [Event Log] \label{def:event_log}
An {\em event log} (denoted by $\mathcal{L}$) is a set of executed activity instances, $E \subseteq \mathcal{E}$, 
with each event having a schema $\sigma_\mathcal{E} = \{\xi, \alpha, \tau_{start}, \tau_{end}\}$,
that assigns the following attribute values to events: 
\begin{compactitem}
\item $\xi: \mathcal{E} \rightarrow C$ assigns a case identifier,
\item $\alpha:  \mathcal{E} \rightarrow A$ assigns an activity label, and,
\item $\tau_{start}, \tau_{end}: \mathcal{E} \rightarrow T$ assign the start and end timestamps, respectively. 
\end{compactitem}
\end{defn} 

\begin{table}[t]
	\centering \scriptsize
    \caption{Example of 6 events of an event log from a Procure-to-Pay Process}
    \label{tab:example_log}
    
    \setlength{\tabcolsep}{3pt}
	\begin{tabular}{c l r r l | c l r r l}
        \toprule
        Case &  \multicolumn{1}{c}{Activity}       &  \multicolumn{1}{c}{Start}     & \multicolumn{1}{c}{End}        &&& Case & \multicolumn{1}{c}{Activity}      & \multicolumn{1}{c}{Start}      &  \multicolumn{1}{c}{End}       \\ \toprule \toprule
        111  &  CreatePO       &   7:12:00  &   7:20:00  &&&  222 & CreatePO      &  10:12:00  &  10:47:00  \\
        111  &  ApprovePO      &   9:30:00  &  10:12:00  &&&  222 & PO\_Rejected  &  10:47:00  &  11:26:00  \\
        111  &  GoodsReceived  &  10:12:00  &  10:44:00  &&&  333 & CreatePO      &   9:26:00  &  10:32:00  \\ \bottomrule
    \end{tabular}
    \vspace*{-3mm}
\end{table}

\subsection{Measures for Time-Series and Histogram Comparison}


To analyze the temporal performance of a process, an event log can be mapped to a variety of time series (e.g.\ activity starts, activity ends). 
Accordingly, we consider the use of techniques to quantity the distance between two time series, $x = (x_1, \ldots, x_n)$, and $y = (y_1, \ldots, y_m)$,
of (potentially different) lengths $n$ and $m$, respectively.
 To this end, 
one may employ various measures, such as computing $||x-y||_{l}$ in any of the standard norms (i.e., $l=1, 2, \infty$).\footnote{See Section 2.2 in~\cite{DBLP:journals/pr/Liao05} for a survey on time series comparison measures.}
These comparisons would only
be possible after padding the shorter time-series. In addition,
standard norms do not capture the temporal differences
between the two time-series. For example, a temporal shift in $x$ vs $y$ may produce $l1$ or $l2$ norm, but represents a significant failure in the model to capture time-series pattern properly. To overcome the two limitations, 
namely the need for padding, and ignoring temporal differences, a 
natural measure 
is the Wasserstein Distance (WD)~\cite{muskulus2011wasserstein};
in this work, we consider two variations of WD. 
\begin{itemize}
\item 
\emph{Earth Mover's Distance (EMD)}~\cite{DBLP:conf/bpm/LeemansSA19} 
computes the effort it takes to balance two vectors $x$ and $y$ of different lengths, treating each entry $x_i, y_j$ as `masses' 
to move from location to location until the two time-series are equal. EMD does not assume
that \begin{math} \sum_i x_i = \sum_j y_j \end{math},
i.e., the sum of the `earth mass' to be moved 
can be different; in such cases, we add a penalty for creating redundant mass to 
fill in gaps. Herein, we consider the EMD problem 
with absolute distance measure~\cite{DBLP:conf/iccv/LevinaB01}.
\item
\emph{1st Wasserstein Distance (1WD)}~\cite{DBLP:conf/iccv/LevinaB01} is a computationally efficient 
variation of the EMD. It
introduces the constraint that the sum of masses must be
the same in $x$ and $y$ 
(i.e., the constraint \begin{math} \sum_i x_i = \sum_j y_j \end{math} is enforced).
1WD is suitable for comparing
empirical distribution functions (histograms), 
since the sum of the mass in each is $1$. 
\end{itemize}

When comparing two histograms, we let $f = (f_1, \ldots, f_n) $ be the $n$ normalized frequency values of the first histogram,
and let $g = (g_1, \ldots, g_m) $ be the $m$ normalized frequencies of the second histogram.
We treat the two histograms $f$ and $g$ similarly to the two time series 
$x$ and $y$,
and employ 1WD distance, since the sum of masses is $1$ (EMD and 1WD lead to the same results).

\section{Framework for Measuring BPS Model Quality\label{sec:approach}}

In this part, we develop an approach for measuring the quality of BPS models.
There are two main reasons why directly evaluating a BPS model would be impractical:
\textit{i)} typically, the `true' BPS model of the process is not available (and often does not exist), thus, we cannot perform a model-to-model comparison;
\textit{ii)} different simulation engines (e.g., Prosimos~\cite{DBLP:conf/bpm/Lopez-PintadoD22}) support different BPS model formats, hindering a generic comparison of BPS models.
Therefore, we propose to generate a collection of logs simulated with the BPS model under evaluation, and compare them to event logs of the actual system (i.e., the system that the model aims to mimic).
Consequently, one can apply a `transitive argument': the `closer' the simulated logs are to the actual data, the better is the model.
In other words, we treat the (test) data as our `ground truth', since useful models are supposed to be faithful generators of `reality'.

Two challenges arise when measuring the quality of a BPS model:
\textit{i)} a model can be very close to the data in one aspect (e.g., control-flow), yet very different in another (e.g., in inter-arrival times), and, 
\textit{ii)} a model can generate many realities as it is probabilistic in nature (durations and routing are stochastic), while the data consists of a single realization.
To overcome the first limitation, we propose a collection of measures to quantify the distance across multiple process perspectives.
Specifically, we shall consider control-flow, temporal, and congestion distance measures.
As for the second limitation, our approach is to generate multiple event logs simulating the `ground-truth' event log (i.e., with the same number of cases, and starting from the same instant in time), and use the generated logs to construct confidence intervals for each of our measures.

For all measures, we consider a collection of $K$ generated logs (GLogs) that came from $K$ simulation runs, and compare these $K$ GLogs to the actual test event log (ALog) that, importantly, was not used to construct the BPS model.
Below, we outline control-flow, temporal, and congestion, discuss their rationale, and briefly provide their computation by comparing GLogs and an ALog.\footnote{
For clarity, each measure is described as a distance between two event logs.
However, we propose to report the average of $K$ individual comparisons (GLog against ALog).}


\subsection{Control-flow Measures}

To evaluate the quality w.r.t.\ the control-flow perspective (i.e., the capability of the model to represent the event sequences in the actual event log), we propose two measures.
The first one, namely control-flow log distance (CFLD), is a variation of a measure introduced by Camargo et al. in~\cite{DBLP:journals/peerj-cs/CamargoDR21}.
CFLD precisely penalizes the differences in the control-flow by pairing each case in GLog to the case in the ALog that minimizes the sum of their distances.
However, due to its steep computational complexity, we propose an additional measure, the n-gram distance (NGD), that approaches the problem in a more efficient way.


\medskip\noindent\textbf{Control-flow Log Distance (CFLD).}
Given two logs $\mathcal{L}_1$ and $\mathcal{L}_2$ with the same number of cases, we compute
the average distance to transform each case in $\mathcal{L}_1$ to another case in $\mathcal{L}_2$ (see~\cite{DBLP:journals/peerj-cs/CamargoDR21} for a description of a similarity version of this measure).
We transform each process case of $\mathcal{L}_1$ and $\mathcal{L}_2$ to their corresponding activity sequences, abstracting from temporal information.
Then, we compute the Damerau-Levenshtein (DL) distance~\cite{DBLP:journals/bmcbi/ZhaoS19} between each pair of cases $i, j$ belonging to $\mathcal{L}_1$ and $\mathcal{L}_2$, respectively, normalizing them by the maximum of their lengths (obtaining a value in $[0, 1]$).
Subsequently, we compute the matching between the cases of both logs (such that each $i$ is matched to a different $j$, and vice versa) minimizing the sum of distances using the Hungarian algorithm for optimal alignment. The CFLD is the average of the normalized distance values.

CFLD requires pairing each case in the simulated log with a case in the original log, minimizing the total sum of distances. The computational complexity of computing the DL-distance for all possible pairings is $O(N^2 \times MTL^3)$ where N is the number of traces in the logs (assuming both logs have an equal number of cases, which holds in our setting) and $MTL$ is the maximum trace length. Since all pairings are put into a matrix to compute the optimal alignment of cases (the one that minimizes the total sum of distances), CFLD's memory complexity is quadratic on the number of cases. The optimal alignment of traces using the Hungarian algorithm has a cubic complexity on the number of cases. 

\medskip\noindent\textbf{N-Gram Distance (NGD).} 
Leemans et al.~\cite{DBLP:conf/bpm/LeemansSA19} measure the quality of a stochastic process model by mapping the model and a log to their Directly-Follows Graph (DFG), viewing each DFG as a histogram, and measuring the distance between these histograms. We note that the histogram of 2-grams of a log is equal to the histogram of its DFG.\footnote{An n-gram is a vector of $n$ consecutive activities in a trace of a log. A 2-gram is a pair of consecutive activities in a log. Every arc in the DFG of a log is a 2-gram of the log and vice-versa.} Given this observation, we generalize the approach of~\cite{DBLP:conf/bpm/LeemansSA19} to n-grams, noting that the histogram of n-grams of a log is equal to the (n-1)$^{\mathrm{th}}$-Markovian abstraction of the log~\cite{DBLP:journals/tkde/AugustoACDR22}. In other words, the histogram of 2-grams is the $1^{st}$-order Markovian abstraction (the DFG), the histogram of 2-grams is the $2^{nd}$-order Markovian abstraction, and so on. 

Given two logs $\mathcal{L}_1$ and $\mathcal{L}_2$, and a positive integer $n$, we compute the difference in the frequencies of the n-grams observed in $\mathcal{L}_1 \bigcup \mathcal{L}_2$.
To compute this measure, we transform each case of $\mathcal{L}_1$ and $\mathcal{L}_2$ to its corresponding activity sequences, abstracting temporal information, and adding $n-1$ dummy activities to both start and end of the case (e.g., \texttt{0-A-B-C-0} for case \texttt{A-B-C} and $n=2$).
Then, we compute all sequences of 
$n$ activities (n-grams) observed in 
each log, and measure their frequency.
Finally, we compute the sum of absolute differences between the frequencies of each computed n-gram, and normalize the total distance by the sum of frequencies of all n-grams in both logs (obtaining a value in $[0, 1]$).

For example, consider $\mathcal{L}_1$ having three cases \texttt{A-B-C-D}, and $\mathcal{L}_2$ having three cases \texttt{A-B-E-D}.
Given $n=2$, the observed n-grams are \texttt{0-A}, \texttt{A-B}, \texttt{B-C}, \texttt{C-D}, and \texttt{D-0} in $\mathcal{L}_1$; and \texttt{0-A}, \texttt{A-B}, \texttt{B-E}, \texttt{E-D}, and \texttt{D-0} in $\mathcal{L}_2$ (each one with a frequency of three).
The n-grams \texttt{B-C}, \texttt{C-D}, \texttt{B-E}, and \texttt{E-D} have a frequency of 3 in one log, and 0 in the other, thus, the NGD between $\mathcal{L}_1$ and $\mathcal{L}_2$ is 0.4 (12 divided by 30).
By adding dummy activities, all activity instances have the same weight in the measure, as each of them is present in $n$ n-grams.
Otherwise, the first and last activity instances of each trace would be present only in one n-gram.
Note that we do not use the EMD to compute the NGD, because the order of the n-grams in the histogram is irrelevant and EMD would take this order into account.

NGD is considerably more efficient than  CFLD, as the construction of the histogram of n-grams is linear on the number of events in the log, and the same goes for computing the differences between the n-gram histograms.

\subsection{Temporal Measures}

We propose three measures that assess the ability of a BPS model to capture the temporal performance perspective, based on the idea that the time series of events generated by a BPS model should be similar to the time series of the test data, with respect to seasonality, trend, and time-to-event.

The first two measures come from time-series analysis, where most approaches in the literature (e.g.,\ SARIMA) decompose the time series into components of trend, seasonality, and noise~\cite{brockwell2002introduction}.
We follow a similar path by analyzing the trend (comparing the absolute distribution of events), and the seasonality (comparing the circadian distribution of events).
The third measure comes from time-to-event (or survival) analysis~\cite{kalbfleisch2011statistical}, 
a field in statistics that analyzes the behavior of individuals from some point in time until 
an event of interest occurs. 
Specifically, we are interested in analyzing the capability of the simulator to correctly reconstruct the occurrence of events (and their timestamps) from the beginning of the corresponding 
case to its end. 


\medskip\noindent\textbf{Absolute Event Distribution (AED).} 
Given two event logs $\mathcal{L}_1$ and $\mathcal{L}_2$, we transform the events into a time series by binning the timestamps in the event log (both start and end) by date and hour of the day (e.g., timestamps between `02/05/2022 10:00:00' and `02/05/2022 10:59:59' will be placed into the same bin). 
Let $i = 1, \ldots, B$ be the hours from the first until the last timestamp in $\mathcal{L}_1 \bigcup \mathcal{L}_2$ (i.e., the timeline of both logs), and $dh(\tau(e))$ a function returning the $i$ corresponding to the date and hour of the day of a timestamp of event $e$ (for brevity, we refer to both $\tau_{start}$ and $\tau_{end}$ as $\tau$), the binning procedure is as follows,
\begin{equation} \label{eq:binning}
    x_i = \lvert\{ e \in \mathcal{L}_1 \mid dh(\tau(e)) = i\}\rvert , \ \ \ 
    y_i = \lvert\{ e \in \mathcal{L}_2 \mid dh(\tau(e)) = i\}\rvert
\end{equation}
Finally, the AED distance between $\mathcal{L}_1$ and $\mathcal{L}_2$ corresponds to the EMD between $x_1, \ldots, x_B$ and $y_1, \ldots, y_B$.

\medskip\noindent\textbf{Circadian Event Distribution (CED).} 
Given two event logs $\mathcal{L}_1$ and $\mathcal{L}_2$, we partition each log 
into sub-logs by the day of the week (Mon-Sun).
Let $wd(\tau(e))$ be a function that returns the day of the week for timestamp $\tau(e)$.
Then, for $i = 1, \ldots, 7$, we obtain the corresponding sub-logs as follows, 
\begin{equation}
    \mathcal{L}_{1, i}= \{ e \in \mathcal{L}_1 \ | \ wd(\tau(e)) = i\}, \ \ \ 
    \mathcal{L}_{2, i}= \{ e \in \mathcal{L}_2 \ | \ wd(\tau(e)) = i\}
\end{equation}
Subsequently, we bin each sub-log into hours with Eq.~\eqref{eq:binning} using $h(\tau(e))$, a function returning the hour of the day of a timestamp of event $e$, instead of $dh(\tau(e))$.
In this way, all the timestamps recorded on any Monday between `10:00:00' and `10:59:59' will be placed in the same bin), obtaining $x_{1, d}, \ldots, x_{B, d}$ and $y_{1, d}, \ldots, y_{B, d}$ with $d \in \{1, \ldots, 7\}$.
Finally, the CED distance between $\mathcal{L}_1$ and $\mathcal{L}_2$ corresponds to the average of the EMD between $x_{1, d}, \ldots, x_{B, d}$ and $y_{1, d}, \ldots, y_{B, d}$ with $d \in \{1, \ldots, 7\}$.


\medskip\noindent\textbf{Relative Event Distribution (RED).} 
Here, we wish to analyze the ability of the simulator to mimic the temporal distribution of events w.r.t.\ the origin of the case (i.e., the case arrival).
To this end, given two event logs $\mathcal{L}_1$ and $\mathcal{L}_2$, we offset all log timestamps from their corresponding case arrival time (the first timestamp in a case is set to time $0$, the second one is set to the inter-event time from the first, etc.).
Formally, let $a(\xi(e)) = \min_{t'} \{ t' \ | \ t' = \tau_{start}(e') \wedge e' \in \mathcal{L} \wedge \xi(e') = \xi(e)\} $ be the arrival time of
a case associated with an event in the log. Then, the relative event times $\rho(e)$ are defined as,
\begin{equation} \label{eq:relative}
    \rho(e) = \tau(e) - a(\xi(e)),
\end{equation}
\vspace*{-5mm}

with $\tau(e)$ being $\tau_{start}(e)$ for start times, and $\tau_{end}(e)$ denoting end times.
We apply Eq.~\eqref{eq:relative} to the timestamps in $\mathcal{L}_1$ and $\mathcal{L}_2$ and, for each log, discretize the resulting $\rho(e)$ into hourly bins (e.g., those durations between 0 and 3,599 seconds go to the same bin).
Finally, the RED distance between $\mathcal{L}_1$ and $\mathcal{L}_2$ corresponds to the EMD between the discretized $\rho(e)$ of each log.

\subsection{Congestion Measures}

To measure the capability of a model to represent congestion, we rely on queueing theory, a field in applied probability that 
studies the behavior of congested systems~\cite{thomas1976queueing}. 
The workload
in a queueing system is dominated by two factors: the \textit{arrival rate} of
cases over time, and the \textit{cycle time}, which is the length-of-stay of a case in the system. 
Below, we propose two measures to compare the two workload components over pairs of event logs
by comparing the time series of the arrivals, and the distribution 
of the cycle times (assuming that its variability is captured by the arrivals time-series comparison).


\medskip\noindent\textbf{Case Arrival Rate (CAR).}
This measure compares case arrival patterns (shape) and counts (number of arrival per bin).
Given two event logs $\mathcal{L}_1$ and $\mathcal{L}_2$, we use the function $a(c), c \in C$ to obtain the sets of arrival timestamps of each log.
Subsequently, we bin them using Eq.~\eqref{eq:binning} (timestamps between `02/05/2022 10:00:00` and `02/05/2022 10:59:59` are placed in the same bin), obtaining two vectors $x_1, \ldots, x_B$ and $y_1, \ldots, y_B$ corresponding to the binned arrival timestamps of $\mathcal{L}_1$ and $\mathcal{L}_2$, respectively.
Finally, the CAR distance between $\mathcal{L}_1$ and $\mathcal{L}_2$ corresponds to the EMD between $x_1, \ldots, x_B$ and $y_1, \ldots, y_B$.

\medskip\noindent\textbf{Cycle Time Distribution (CTD).}
Here, we seek to measure the ability of the BPS model to capture the end-to-end cycle time of the process. Given two event logs $\mathcal{L}_1$ and $\mathcal{L}_2$, we collect all cycle times into a single histogram per log, which depicts their empirical probability distribution functions (PDF).
The CTD distance between $\mathcal{L}_1$ and $\mathcal{L}_2$ corresponds to the 1WD between both histograms.\footnote{Since we are comparing two distributions, 1WD and EMD yield the same result.}

\section{Evaluation\label{sec:evaluation}}

We report on a two-fold experimental evaluation.
The first part aims to validate the applicability of the proposed measures by testing the following evaluation question: \textit{are the proposed measures able to discern the impact of different known modifications to a BPS model?} \textbf{(EQ1)}.
Given the potential efficiency issues of CFLD, the first part of the evaluation also aims to answer the question: \textit{is the N-Gram Distance's performance significantly different from the CFLD's performance?} \textbf{(EQ2)}.
The second part of the evaluation is designed to test if: \textit{given two BPS models discovered by existing automated BPS model discovery techniques in real-life scenarios, are the proposed measures able to identify the strengths and weaknesses of each technique?} \textbf{(EQ3)}.
Given the complexity of the EMD (cf.\ \sectionname~\ref{sec:preliminaries}), the second part of this evaluation also focuses on answering: \textit{does the {1-WD} report the same insights in real-life scenarios as the EMD?} \textbf{(EQ4)}. 

In the case of the NGD, we report on this measure for a size $N=2$.\footnote{
Augusto et al.~\cite{DBLP:journals/tkde/AugustoACDR22} found that, for models with no duplicate activities, the size of $N=2$ captures enough information to compare processes from a control-flow perspective.
}
The distance computed by the EMD is not directly interpretable, as it is an absolute number on a scale that depends on the range of values of the input time series.
Accordingly, we divide the raw EMD by the number of observations in the original log. 
In this way, we can interpret the resulting scaled-down EMD as the average number of bins that each observation of the original log must be moved to transform it into the simulated log.
For example, a value of 10 implies that, on average, each observation had to be moved 10 bins. 

\subsection{Synthetic Evaluation\label{subsec:evaluation-synthetic}}

\noindent \textbf{Datasets.}
To assess EQ1 and EQ2, we manually created the BPS model of a loan application process based on the examples from~\cite[Chapter 10.8]{DBLP:books/sp/DumasRMR18}.
The process comprises 12 activities 
(with one loop, a 3-branch parallel structure, 3 exclusive split gateways, and 3 possible endings) and 6 different resource types.
We simulated a log of 1,000 cases as the log recording the process (i.e., the ALog).
We created 7 modifications of the original BPS model:
\textit{i)} altering the control-flow by arranging the parallel activities as a sequence ($\text{Loan}_{SEQ}$);
\textit{ii)} altering, on top of the previous modification, the branching probabilities ($\text{Loan}_{S \mhyphen G}$);
\textit{iii)} modifying the rate of case arrivals ($\text{Loan}_{ARR}$);
\textit{iv)} increasing the duration of the activities of the process ($\text{Loan}_{DUR}$);
\textit{v)} halving the available resources to create resource contention ($\text{Loan}_{RC}$);
\textit{vi)} changing the resource working schedules from 9am-5pm to 2pm-10pm ($\text{Loan}_{CAL}$); and
\textit{vii)} adding timer events to simulate extraneous waiting time~\cite{DBLP:conf/icpm/Chapela-CampaD22} delaying the start of 4 of the activities ($\text{Loan}_{EXT}$).

We simulated $K=10$ logs (as the GLogs) with 1,000 cases for each altered BPS model. 
\tablename~\ref{tab:synthetic-evaluation} shows the results of the proposed measures for each modified scenario, and for the original BPS model as ground truth ($\text{Loan}_{GT}$) to measure the distance associated with the stochastic nature of the simulation.


\begin{table}[t]
    \centering \scriptsize
    \caption{
      Results (average and 95\% confidence interval) of the proposed measures for the original and modified BPS models of a loan application process.
    }
    \label{tab:synthetic-evaluation}
    
    \setlength\tabcolsep{0pt}
    \begin{tabular}{L{2cm} R{2cm} R{0.5cm} R{2cm} R{0.5cm} R{2cm} R{0.5cm} R{2cm}}
        \toprule            
                                      &  \multicolumn{1}{c}{NGD}       &&  \multicolumn{1}{c}{CFLD}      &&  \multicolumn{1}{c}{AED}                &&  \multicolumn{1}{c}{CED}    \\ \toprule \toprule
                                      
        $\text{Loan}_{GT}$            &  \colorcell0.02 ($\pm$0.00)    &&  \colorcell0.02 ($\pm$0.00)    &&  \colorcell2.39 ($\pm$\phantom{1}0.51)  &&  \colorcell0.05 ($\pm$0.01) \\ 
        $\text{Loan}_{SEQ}$           &  0.34 ($\pm$0.00)              &&  0.20 ($\pm$0.00)              &&  \colorcell2.58 ($\pm$\phantom{1}0.34)  &&  \colorcell0.05 ($\pm$0.01) \\ 
        $\text{Loan}_{S \mhyphen G}$  &  0.46 ($\pm$0.00)              &&  0.32 ($\pm$0.00)              &&  44.17 ($\pm$\phantom{1}9.59)           &&  0.24 ($\pm$0.01)           \\
        $\text{Loan}_{ARR}$           &  0.04 ($\pm$0.01)              &&  0.03 ($\pm$0.00)              &&  35.66 ($\pm$14.46)                     &&  0.08 ($\pm$0.01)           \\
        $\text{Loan}_{DUR}$           &  \colorcell0.03 ($\pm$0.01)    &&  \colorcell0.02 ($\pm$0.00)    &&  4.29 ($\pm$\phantom{1}1.34)            &&  \colorcell0.05 ($\pm$0.01) \\ 
        $\text{Loan}_{RC}$            &  0.23 ($\pm$0.03)              &&  0.15 ($\pm$0.01)              &&  23.67 ($\pm$\phantom{1}8.81)           &&  \colorcell0.06 ($\pm$0.01) \\
        $\text{Loan}_{CAL}$           &  \colorcell0.02 ($\pm$0.00)    &&  \colorcell0.02 ($\pm$0.00)    &&  6.27 ($\pm$\phantom{1}0.37)            &&  3.51 ($\pm$0.02)           \\ 
        $\text{Loan}_{EXT}$           &  \colorcell0.02 ($\pm$0.01)    &&  \colorcell0.02 ($\pm$0.00)    &&  5.43 ($\pm$\phantom{1}1.64)            &&  0.09 ($\pm$0.01)           \\
    \end{tabular}
    \begin{tabular}{L{2cm} R{2.5cm} R{1cm} R{2.5cm} R{1cm} R{2.5cm}}
        \toprule
                                      &  \multicolumn{1}{c}{RED}                &&  \multicolumn{1}{c}{CAR}                &&  \multicolumn{1}{c}{CTD}                \\ \toprule \toprule
                                      
        $\text{Loan}_{GT}$            &  \colorcell0.22 ($\pm$\phantom{1}0.05)  &&  \colorcell0.00 ($\pm$\phantom{1}0.00)  &&  \colorcell7.06 ($\pm$\phantom{11}1.41) \\ 
        $\text{Loan}_{SEQ}$           &  1.91 ($\pm$\phantom{1}0.26)            &&  \colorcell0.00 ($\pm$\phantom{1}0.00)  &&  42.38 ($\pm$\phantom{11}3.83)          \\ 
        $\text{Loan}_{S \mhyphen G}$  &  235.36 ($\pm$12.93)                    &&  \colorcell0.00 ($\pm$\phantom{1}0.00)  &&  7,667.13 ($\pm$443.38)                 \\
        $\text{Loan}_{ARR}$           &  0.62 ($\pm$\phantom{1}0.16)            &&  42.39 ($\pm$14.12)                     &&  11.89 ($\pm$\phantom{11}2.42)          \\
        $\text{Loan}_{DUR}$           &  7.09 ($\pm$\phantom{1}0.51)            &&  0.09 ($\pm$\phantom{1}0.04)            &&  200.53 ($\pm$\phantom{1}15.66)         \\ 
        $\text{Loan}_{RC}$            &  31.66 ($\pm$\phantom{1}8.62)           &&  0.03 ($\pm$\phantom{1}0.02)            &&  759.45 ($\pm$210.93)                   \\
        $\text{Loan}_{CAL}$           &  \colorcell0.26 ($\pm$\phantom{1}0.06)  &&  6.24 ($\pm$\phantom{1}0.00)            &&  \colorcell7.51 ($\pm$\phantom{11}1.94) \\ 
        $\text{Loan}_{EXT}$           &  8.02 ($\pm$\phantom{1}0.18)            &&  \colorcell0.00 ($\pm$\phantom{1}0.00)  &&  262.30 ($\pm$\phantom{11}6.19)         \\ \bottomrule
    \end{tabular}
    \vspace*{-3mm}
\end{table}

\medskip
\noindent \textbf{Results and Discussion.}
Regarding EQ1, \tablename~\ref{tab:synthetic-evaluation} shows how the proposed measures appropriately penalize the BPS models for the modifications affecting their corresponding perspectives.
In the control-flow measures, the BPS models showing significant differences w.r.t.\ the ground truth are those with control-flow modifications.
The distances of $\text{Loan}_{RC}$ and $\text{Loan}_{SEQ}$ are explained by the parallel activities being executed more frequently in a specific order.
In the first BPS model, due to resource contention, which delays the execution of one of the parallel activities in some cases.
In the second one, due to the control-flow modification.
Finally, $\text{Loan}_{S \mhyphen G}$ reports the highest distance as, in addition to the modification in $\text{Loan}_{SEQ}$, it also alters the frequency of each process variant.

For temporal measures, the AED distance captures 
the difference in the distribution of events along the entire process.
However, to identify the sources of these differences.
we require a combination of the penalties incurred by CED, RED, and CAR. 
Thus, we must analyze them to find the root-causes for the discrepancies in AED. 
Starting from the seasonal aspects captured by 
CED, only $\text{Loan}_{S \mhyphen G}$ and $\text{Loan}_{CAL}$ 
report significant differences, being the latter the only BPS model altering seasonal aspects.
$\text{Loan}_{S \mhyphen G}$'s distance is due to the change in the gateway probabilities,
which in turns impacts the overall distribution of executed events.
As expected, $\text{Loan}_{CAL}$ presents the highest CED distance due to the change in schedules 
that displaces executed events from morning to evening.

Moving to RED, which reports the distance in the distribution of events over time within each case,
we observe that all modifications except $\text{Loan}_{CAL}$ should affect this perspective.
The slightly higher penalization of $\text{Loan}_{ARR}$ is due to the 
higher case arrival rates, which delay the start activities due to resource contention.
$\text{Loan}_{SEQ}$ presents a higher distance 
(close to a displacement of 2 hours per event) 
as the three parallel activities are executed as a sequence, delaying subsequent activities.
Similarly, in $\text{Loan}_{DUR}$, $\text{Loan}_{EXT}$, and $\text{Loan}_{RC}$, activity delays are caused by longer durations, extraneous delays, and resource contention waiting times, respectively.
Finally, $\text{Loan}_{S \mhyphen G}$ presents the highest RED distance due to the high frequency differences in each process variant.

Switching to CAR, we do not observe significant differences in BPS models 
that exhibit the same arrival rate, except for $\text{Loan}_{CAL}$.
The latter is explained by the change in schedules, as cases 
cannot start until the resources start their working period (which skews effective start times).
Unsurprisingly, for $\text{Loan}_{ARR}$, the difference in CAR is due to the change in the arrival model.

Finally, the last proposed measure is CTD, which reports the distance in case duration among all the cases.
The results of CTD 
follow a similar to RED (yet, with different values), 
since cycle times correspond to the time distance between the first and last events of the case.
However, this correlation might not hold across all scenarios.
Specifically, if the distribution of executed activities in the 
middle of each case is different, but the last event does not change, RED would detect discrepancies that CTD would not 
(as the cycle time would remain the same). Thus, CTD is 
most relevant when the analysis revolves around total cycle times, 
while disregarding the temporal distribution of events within the case.

To answer EQ2, we computed the Kendall rank correlation coefficient between NGD and CFLD, and we obtained a correlation of 1.0. Thus, in light of the complexity of CFLD (cf. Sect.~\ref{sec:approach}), we recommend using NGD to assess the quality of a BPS model from the control-flow perspective.

\subsection{Real-life Evaluation\label{subsec:evaluation-real-life}}

\noindent \textbf{Datasets.}
To evaluate EQ3 and EQ4, we selected four real-life logs of different complexities: \textit{i)} a log from an academic credentials' management process (AC\_CRE), containing a high number of resources exhibiting low participation in the process. \textit{ii)} a log of a loan application process from the Business Process Intelligence Challenge (BPIC) of 2012\footnote{\url{https://doi.org/10.4121/uuid:3926db30-f712-4394-aebc-75976070e91f}} -- we preprocessed this log by retaining only the events corresponding to activities performed by human resources (i.e., only activity instances that have a duration). \textit{iii)} the log from the BPIC of 2017\footnote{\url{https://doi.org/10.4121/uuid:5f3067df-f10b-45da-b98b-86ae4c7a310b}} -- we pre-processed this log by following the recommendations reported by the winning teams participating in the competition.\footnote{\url{https://www.win.tue.nl/bpi/doku.php?id=2017:challenge}} And \textit{iv)} a log from a call centre process (CALL) containing numerous cases of short duration -- on
average, two activities per case.
To avoid data leakage, we split the log of each dataset into two sets (\textit{training} and \textit{testing}).
These datasets correspond to disjoint (non-overlapping) intervals in time with similar case and event intensity. The training dataset contains cases that are fully contained in the training period, and same for the testing dataset.
\tablename~\ref{tab:log-characteristics} shows the characteristics of the four training and four testing datasets. For each dataset, we ran two automated BPS model discovery techniques (SIMOD and ServiceMiner) on the training log, and evaluated the quality of the discovered BPS models on the test log.

\begin{table}[t]
    \centering \scriptsize
    \caption{Characteristics of the real-life logs used in the evaluation.}
    \label{tab:log-characteristics}
    
    \setlength{\tabcolsep}{9pt}
    \begin{tabular}{l r r r r r}
        \toprule
                     \multicolumn{1}{c}{Event log} & \multicolumn{1}{c}{Cases} & \multicolumn{1}{c}{Activity instances} & \multicolumn{1}{c}{Variants} & \multicolumn{1}{c}{Activities} & \multicolumn{1}{c}{Resources} \\ \toprule \toprule
        AC\_CRE\_TR  & 398      & 1,945     & 54      & 16  & 306     \\
        AC\_CRE\_TE  & 398      & 1,788     & 35      & 16  & 281     \\ \midrule
        BPIC12\_TR   & 3,030    & 16,338    & 735     & 6   & 47      \\
        BPIC12\_TE   & 2,976    & 18,568    & 868     & 6   & 53      \\ \midrule
        BPIC17\_TR   & 7,402    & 53,332    & 1,843   & 7   & 105     \\
        BPIC17\_TE   & 7,376    & 52,010    & 1,830   & 7   & 113     \\ \midrule
        CALL\_TR     & 260,889  & 445,567   & 1,689   & 19  & 2,712   \\
        CALL\_TE     & 260,890  & 454,807   & 1,573   & 19  & 2,711   \\ \bottomrule
    \end{tabular}
    \vspace*{-3mm}
\end{table}

\begin{table}[t]
  \centering \scriptsize
  \caption{
    Distance measures for the BPS models discovered by SIMOD and ServiceMiner on the logs in \tablename~\ref{tab:log-characteristics}.
    The CFLD ran out of memory (48 GB of allocated memory) on the CALL dataset after $>$ 2 hours, thus no values are reported in those cells.
  }
  \label{tab:real-life-evaluation}
  
  \setlength{\tabcolsep}{2pt}
  \begin{tabular}{L{1cm} L{2cm} r r r r r r r}
        \toprule
                            &                &  \multicolumn{1}{c}{AC\_CRE}              &&  \multicolumn{1}{c}{BPIC12}             &&  \multicolumn{1}{c}{BPIC17}      &&  \multicolumn{1}{c}{CALL}     \\ \toprule \toprule
                            
    \multirow{2}{*}{NGD}    &  SIMOD         &  0.24 ($\pm$0.01)                         &&  0.56 ($\pm$0.00)                       &&  0.37 ($\pm$0.00)                &&  0.08 ($\pm$0.00)             \\ 
                            &  ServiceMiner  &  \colorcell0.13 ($\pm$0.01)               &&  \colorcell0.13 ($\pm$0.01)             &&  \colorcell0.06 ($\pm$0.00)      && \colorcell0.04  ($\pm$0.00)   \\ \midrule
    \multirow{2}{*}{CFLD}   &  SIMOD         &  0.21 ($\pm$0.00)                         &&  0.55 ($\pm$0.00)                       &&  0.34 ($\pm$0.00)                &&  -                            \\ 
                            &  ServiceMiner  & \colorcell0.18  ($\pm$0.01)               &&  \colorcell0.16 ($\pm$0.00)             &&  \colorcell0.06 ($\pm$0.00)      &&  -                            \\ \midrule
    \multirow{2}{*}{AED}    &  SIMOD         &  \colorcell91.72 ($\pm$\phantom{1}16.66)  &&  61.57 ($\pm$10.80)                     &&  192.18 ($\pm$33.03)             &&  48.13 ($\pm$0.11)            \\ 
                            &  ServiceMiner  &  298.17 ($\pm$105.87)                     &&  \colorcell29.22 ($\pm$\phantom{1}8.33) &&  \colorcell51.24 ($\pm$11.90)    && \colorcell1.67 ($\pm$0.13)    \\ \midrule
    \multirow{2}{*}{CAR}    &  SIMOD         &  \colorcell110.38 ($\pm$\phantom{1}16.94) &&  336.42 ($\pm$42.98)                    &&  390.04 ($\pm$43.39)             &&  61.68 ($\pm$0.00)            \\ 
                            &  ServiceMiner  &  327.85 ($\pm 118.33$)                    &&  \colorcell153.25 ($\pm$24.76)          &&  \colorcell121.90  ($\pm$23.60)  &&  \colorcell3.57 ($\pm$0.20)   \\ \midrule
    \multirow{2}{*}{CED}    &  SIMOD         &  2.22 ($\pm$0.09)                         &&  \colorcell20.55 ($\pm$0.82)            &&  10.72 ($\pm$0.50)               &&  18.17 ($\pm$0.04)            \\ 
                            &  ServiceMiner  &  \colorcell1.63 ($\pm$0.19)               &&  28.52 ($\pm$1.24)                      &&  \colorcell9.87  ($\pm$0.42)     &&  \colorcell0.70 ($\pm$0.01)   \\ \midrule
    \multirow{2}{*}{RED}    &  SIMOD         &  \colorcell9.96 ($\pm$6.46)               &&  \colorcell3.99 ($\pm$0.95)             &&  \colorcell62.80 ($\pm$2.16)     &&  0.10 ($\pm$0.00)             \\ 
                            &  ServiceMiner  &  70.49 ($\pm$4.08)                        &&  45.91 ($\pm$2.09)                      &&  149.60  ($\pm$0.42)             && \colorcell0.03 ($\pm$0.01)    \\ \midrule
    \multirow{2}{*}{CTD}    &  SIMOD         &  \colorcell62.23 ($\pm$1.65)              &&  \colorcell93.45 ($\pm$0.71)            &&  \colorcell102.85 ($\pm$0.84)    &&  \colorcell8.18 ($\pm$0.08)   \\ 
                            &  ServiceMiner  &  99.88 ($\pm$0.30)                        &&  124.21 ($\pm$0.33)                     &&  112.83  ($\pm$0.16)             &&  12.04 ($\pm$0.07)            \\ \bottomrule
  \end{tabular}
    \vspace*{-3mm}
\end{table}

\medskip
\noindent \textbf{Results and Discussion.}
Regarding EQ3, \tablename~\ref{tab:real-life-evaluation} shows the results of the proposed measures for the BPS models automatically discovered by SIMOD and ServiceMiner (henceforth $M_{SIMOD}$ and $M_{SerMin}$, respectively).
From the control-flow perspective, $M_{SerMin}$ 
performs closer to the original log than $M_{SIMOD}$ for all four datasets.
The reason lies in the methods that the approaches use to model the control-flow.
SIMOD is designed to discover an interpretable process model to support modification for \textit{what-if} analyses.
To this end, SIMOD uses a model discovery algorithm that applies 
multiple pruning techniques to simplify the discovered model.
Conversely, ServiceMiner discovers a Markov chain, which yields more accurate results,
yet can lead to complex `spaghetti models'.


Two main differences are reported w.r.t.\ the temporal and congestion aspects.
First, for Case Arrival Rate (CAR), $M_{SerMin}$ presents better results in BPIC12, BPIC17, and CALL, while 
$M_{SIMOD}$ outperforms it in AC\_CRE.
To model the arrival of new cases, ServiceMiner splits the timeline into one-hour windows, and bootstraps the arrivals per time window.
SIMOD computes the inter-arrival times 
(i.e., the time between each arrival and the next one) and estimates a parametrized distribution to model them.
The complexity of ServiceMiner's arrival model allows it to capture better the arrival rate in scenarios where the density of case arrivals per hour is high, and/or the rate of arrivals varies through time (BPIC12, BPIC17, and CALL).
On the contrary, if cases are scattered over time (AC\_CRE), SIMOD's approach presents a better result.

\enlargethispage{\baselineskip}

The second main 
difference lies in the Relative Event Distribution (RED) and the Cycle Time Distribution (CTD) distances.
Here, $M_{SIMOD}$ obtains better results in both measures except in one case.
In the CALL dataset, $M_{SerMin}$ obtains a smaller RED distance (both methods perform well w.r.t.\ the original log).
SIMOD outperforms ServiceMiner due to a high amount of extraneous activity delays (i.e., waiting times not related to the resource allocation or activity performance) exhibited in these processes. Specifically,
SIMOD includes a component to discover extraneous delays, which improves the distribution of the events within the case.
Both techniques perform close to 
the original log in the CALL dataset because extraneous delays are rare for the call centre process.

For seasonality, the Circadian Event Distribution (CED) reports slight differences between 
the two methods for AC\_CRE and BPIC17, and a moderately better result for $M_{SIMOD}$ in BPIC12.
The CALL dataset presents the highest difference, where $M_{SerMin}$ obtains better results,
which can be attributed to its highly accurate arrival model.
The CALL dataset has mostly  
cases with one or two events. 
Hence, case execution depends more on the arrival time of the case, than on the activity performance and 
congestion models.

Combining all the temporal perspectives in one measure, the results of Absolute Event Distance (AED) follow the same distribution as CAR, where $M_{SerMin}$ presents better results in BPIC12, BPIC17, and CALL, while $M_{SIMOD}$ performing better in AC\_CRE.
Although this measure summarizes all the temporal performance in one, it is highly affected by the performance of the arrival model.
A wrong arrival rate propagates the error to all the events per case, displacing them even if their relative distribution is accurate.

The proposed measures detected key differences between the considered BPS model discovery techniques. 
SIMOD's inferior performance in the control-flow perspective is expected, 
given that it takes a simplified process model as input. 
The results highlight the benefits of SIMOD's extraneous waiting time discovery component (a feature that  ServiceMiner does not have). Finally, although ServiceMiner's arrival model achieved the best results in most of the scenarios, the evaluation in AC\_CRE point towards an improvement opportunity in the situation where cases arrive at a slow rate.

\begin{table}[t]
  \centering \scriptsize
  \caption{Results of the proposed measures for the BPS models discovered by SIMOD and ServiceMiner with the real-life logs in \tablename~\ref{tab:log-characteristics}.}
  \label{tab:real-life-evaluation-wass}
    
  \setlength{\tabcolsep}{2pt}
  \begin{tabular}{L{1cm} L{2cm} r r r r r r r}
        \toprule
                            &                &  \multicolumn{1}{c}{AC\_CRE}               &&  \multicolumn{1}{c}{BPIC12}    &&  \multicolumn{1}{c}{BPIC17}    &&  \multicolumn{1}{c}{CALL}        \\ \toprule \toprule
                            
    \multirow{2}{*}{AED}    &  SIMOD         &   \colorcell117.32 ($\pm$\phantom{1}18.85) &&  313.30 ($\pm$41.50)           &&  314.92 ($\pm$43.02)           &&  61.76 ($\pm$0.08)               \\ 
                            &  ServiceMiner  &   315.88 ($\pm$115.60)                     &&  \colorcell79.19 ($\pm$16.12)  &&  \colorcell65.56 ($\pm$12.55)  &&  \colorcell3.47 ($\pm$0.19)      \\ \midrule
    \multirow{2}{*}{CAR}    &  SIMOD         &   \colorcell110.38 ($\pm$\phantom{1}16.94) &&  336.42 ($\pm$42.98)           &&  390.04 ($\pm$43.39)           &&  61.68 ($\pm$0.00)               \\ 
                            &  ServiceMiner  &   327.85 ($\pm$118.33)                     &&  \colorcell153.25 ($\pm$24.76) &&  \colorcell121.89 ($\pm$23.60) &&  \colorcell3.57 ($\pm$0.20)      \\ \midrule
    \multirow{2}{*}{CED}    &  SIMOD         &   3.11 ($\pm$0.18)                         &&  \colorcell2.10 ($\pm$0.07)    &&  \colorcell1.65 ($\pm$0.03)    &&  4.72 ($\pm$0.00)                \\ 
                            &  ServiceMiner  &   \colorcell2.50 ($\pm$0.27)               &&  2.34 ($\pm$0.03)              &&  \colorcell1.65 ($\pm$0.02)    &&  \colorcell1.03 ($\pm$0.03)      \\ \midrule
    \multirow{2}{*}{RED}    &  SIMOD         &   \colorcell48.19 ($\pm$1.72)              &&  \colorcell96.82 ($\pm$1.61)   &&  \colorcell132.31 ($\pm$1.68)  &&  \colorcell0.00 ($\pm$0.00)      \\ 
                            &  ServiceMiner  &   74.50 ($\pm$0.18)                        &&  150.83 ($\pm$0.57)            &&  150.26 ($\pm$0.28)            &&  \colorcell0.00 ($\pm$0.00)      \\ \bottomrule
  \end{tabular}
    \vspace*{-3mm}
\end{table}


To evaluate EQ4, \tablename~\ref{tab:real-life-evaluation-wass} shows the result of AED, CAR, CED, and RED measures when computing the distance with 1WD, instead of EMD.
The results follow the same distribution in all cases, except in the CED measure on the BPIC17 dataset, and the RED measure on the CALL dataset.
In both cases, the slight differences shown by EMD are reduced to a similar value by both techniques.
For arrivals, as explained in \sectionname~\ref{sec:preliminaries}, computing the distance with EMD and 1WD provide 
the same result, as the number of observations in both samples is the same (i.e., the number of cases).
In conclusion, computing the distance using 1WD leads to similar conclusions at a lower computational cost. 
Thus, we recommend using 1WD when the masses of both time series are close to each other, and when the number of observations (amount of mass) is large.

\medskip\noindent \textbf{Threats to Validity.}
This evaluation is potentially affected by the following threats to the validity:
\textit{Internal validity}: the experiments rely only on 8 synthetic BPS models, and 8 automatically discovered BPS models from 4 real-life processes.
The results could be different for other datasets.
\textit{Ecological validity}: the evaluation compares the BPS models against the original log.
While this allows us to measure how well the simulation models replicate the as-is process, it does not allow us to assess the goodness of the simulation models in a what-if setting. 


\section{Conclusion \label{sec:conclusions}}

\enlargethispage{\baselineskip}

We proposed a multi-perspective approach to measure the ability of a BPS model to replicate the behavior recorded in an event log. The approach decomposes simulation models into three perspectives:
control-flow, temporal, and congestion. We defined measures for each of these perspectives. 
We evaluated the adequacy of the  proposed measures by analyzing their ability to discern the impact of modifications to a BPS model.
The results showed that the measures are able to detect the alterations in their corresponding perspectives.
Furthermore, we analyzed the usefulness of the metrics in real-life scenarios w.r.t.\ their ability to uncover the relative strengths and weaknesses of two approaches for automated discovery of BPS models. The findings showed that beyond capturing
the quality of BPS model and identifying the sources of discrepancies,
the measures can also assist in eliciting areas for improvement in these techniques.

In future work, we will explore the applicability of the proposed measures to other process mining problems, e.g., concept drift detection and variant analysis. Studying how to assess the quality of BPS models in the context of object-centric event logs is another future work avenue. Lastly, we aim to study other quality measures for BPS models adapted from the field of generative machine learning.


\medskip\noindent\textbf{Reproducibility.} Scripts to reproduce the experiments, datasets, and results available at: \url{https://doi.org/10.5281/zenodo.7761252}. Implementation at: \url{https://github.com/AutomatedProcessImprovement/log-distance-measures}.

    \bibliographystyle{splncs04}
    \bibliography{references}

\end{document}